\documentstyle[preprint,eqsecnum,aps]{revtex}
\begin{document}
\draft
\title{AMPLITUDE ANALYSIS OF REACTIONS\\
$\pi^- p \to \eta \pi^- p$ AND $\pi^- p\to\eta\pi^0 n$\\
MEASURED ON POLARIZED TARGET\\
AND THE EXOTIC $1^{-+}$ MESON.}
\author{M. Svec\footnote{electronic address: svec@hep.physics.mcgill.ca}}
\address{Physics Department, Dawson College, Montreal, Quebec,  
Canada H3Z 1A4\\
and\\
McGill University, Montreal, Quebec, Canada H3A 2T8}
\maketitle
\begin{abstract}
Recently several experimental groups analysed data on $\pi^- p \to  
\eta \pi^- p$ and $\pi^- p \to \eta \pi^0 n$ reactions with exotic  
$1^{-+}$ $P$-wave and found a conflicting evidence for an exotic  
meson $I=1 1^{-+} (1405)$. High statistics data on these reactions  
are presently analysed by BNL E852 Collaboration. All these analyses  
are based on the crucial assumption that the production amplitudes  
do not depend on nucleon spin. This assumption is in sharp conflict  
with the results of measurements of $\pi^- p \to \pi^- \pi^+ n$,  
$\pi^+ n \to \pi^+ \pi^- p$ and $K^+ n \to K^+ \pi^- p$ on polarized  
targets at CERN which find a strong dependence of production  
amplitudes on nucleon spin. To ascertain the existence of exotic  
meson $1^{-+} (1405)$, it is necessary to perform a  
model-independent amplitude analysis of reactions $\pi^- p \to \eta  
\pi^- p$ and $\pi^- p \to \eta\pi^0 n$. We demonstrate that  
measurements of these reactions on transversely polarized targets  
enable the required model independent amplitude analysis without the  
assumption that production amplitudes are independent on nucleon  
spin. We suggest that high statistics measurements of reactions  
$\pi^- p \to \eta\pi^- p$ and $\pi^- p \to\eta\pi^0 n$ be made on  
polarized targets at BNL and at Protvino IHEP, and that  
model-independent amplitude analyses of this polarized data be  
performed to advance hadron spectroscopy on the level of spin  
dependent production amplitudes.
\end{abstract}
\section{Introduction}

Search for meson states with non-$q\overline q$ quantum numbers  
such as $J^{PC} = 0^{+-}, 1^{-+}, 2^{+-}, \ldots$ has attracted much  
attention in recent years. Of special importance are reactions  
$\pi^- p \to \eta\pi^- p$, $\pi^- p\to\eta\pi^0 n$ and $\pi^-  
p\to\eta\eta^\prime n$. In these reactions the dimeson system is  
produced predominantly in spin states $J = 0$ ($S$-wave), $J=1$  
($P$-wave) and $J=2$ ($D$-wave) for masses up 2.6 GeV. It is the  
$P$-wave which is of special interest as it carries exotic quantum  
numbers $I=1 J^{PC} = 1^{-+}$ for reactions $\pi^- p\to\eta\pi^- p$,  
$\pi^- p\to\eta\pi^0 n$ and $I=0 J^{PC} = 1^{-+}$ for $\pi^- p \to  
\eta\eta^\prime n$.

Measurements of $\pi^- p \to\eta\pi^0 n$ at 100 GeV/c by the GAMS  
Collaboration\cite{alde88} found large Forward-Backward asymmetry  
with pronounced features at around 1300 MeV. Similar  
Forward-Backward asymmetry was found in measurements of $\pi^- p  
\to\eta\pi^- p$ at 6.3 GeV/c by KEK E-179  
Collaboration\cite{aoyagi93,aoyagi94}. The higher statistics  
measurement of $\pi^- p \to\eta\pi^- p$ and $\pi^- p\to\eta\pi^0 n$  
reactions at 18 GeV/c by BNL E-852 Collaboration\cite{coson95}  
confirmed significant Forward-Backward asymmetry in the data  
beginning at invariant mass of about 1.2 GeV in both reactions. The  
behaviour of the asymmetry suggests the presence of large exotic  
$P$-wave interfering with dominant $D$-wave with its $a_2(1320)$  
resonance. The question arises whether there is a resonant  
production of the $\eta\pi^-$ or $\eta\pi^0$ state in the exotic  
$P$-wave. The reliable determination of existence of exotic  
resonance in $1^{-+}$ $P$-wave requires model-independent amplitude  
analysis of the data.

The reactions $\pi^- p\to\eta\pi^- p$, $\pi^- p\to\eta\pi^0 n$ and  
$\pi^- p\to\eta\eta^\prime n$ are described by 14 spin dependent  
production amplitudes: 2 $S$-wave amplitudes $S_n$, 6 $P$-wave  
amplitudes $P^0_n$, $P^-_n$, $P^+_n$ and 6 $D$-wave amplitudes  
$D^0_n$, $D^-_n$, $D^+_n$ where $n=0,1$ is nucleon helicity flip  
$n=|\lambda_p - \lambda_n|$. The amplitudes $S_n$, $P^0_n$, $D^0_n$  
describe the production with dimeson helicity $\lambda=0$ and  
correspond to unnatural exchange. The amplitudes $P^-_n$, $D^-_n$  
and $P^+_n, D^+_n$ describe production with dimeson helicity $\pm 1$  
and correspond to unnatural and natural exchanges, respectively.

All previous amplitude analyses of reactions $\pi^- p \to\eta\pi^-  
p$, $\pi^- p\to\eta\pi^0 n$ and $\pi^- p\to\eta\eta^\prime n$ on  
unpolarized targets are model dependent. They use a very strong  
simplifying assumption that the production amplitudes do not depend  
on nucleon spin\cite{costa80,chung96,chung96b}. The purpose of this  
assumption is to reduce the number of amplitudes by half and thus to  
enable the amplitude analysis of unpolarized moments measured in  
these reactions to proceed. These analyses simply ignore the nucleon  
helicity flip index $n$.

Using such enabling assumption, the different collaborations found  
the exotic $I=1 1^{-+}$ meson but in different amplitudes. The GAMS  
Collaboration reported $1^{-+} (1405)$ state with a width of 180  
MeV\cite{alde88} observed only in the amplitude $|P^0|^2$. The KEK  
E-179 Collaboration\cite{aoyagi93,aoyagi94} found $|P^-|^2$  
nonresonating but found resonance $1^{-+} (1323)$ with a width of  
143 MeV in the amplitude $|P^+|^2$ and possibly in $|P^0|^2$. The  
VES Collaboration\cite{beladid93} measured $\pi^- p\to\eta\pi^- p$  
and $\pi^- p\to\eta^\prime\pi^- p$ at 37 GeV/c at IHEP Protvino, and  
found possible $1^{-+} (1400)$ state only in the amplitude  
$|P^+|^2$. Amplitude analysis of BNL E-852 Collaboration higher  
statistics data at 18 GeV/c is in progress, but it also uses the  
simplifying assumption that production amplitudes do not depend on  
nucleon spin. All these analyses are subjected to an eight-fold  
ambiguity and in Ref.~2, 3 and 8 all eight solutions are presented.

For completeness we note that GAMS Collaboration measured reaction  
$\pi^- p\to\eta\eta^\prime n$ at 38 GeV/c\cite{alde89} and found  
evidence for a new state $X(1920)$. The unusual production and decay  
properties could be understood if $X(1920)$ had a non-$q\overline  
q$ structure, being either a $0^{++}$ or $2^{++}$ glueball or $I=0  
1^{-+}$ exotic meson. Unfortunately, the low statistics does not  
allow even a model dependent amplitude analysis.

The simplifying assumption that the production amplitudes do not  
depend on nucleon spin is not necessary in measurements on polarized  
targets. In 1978, Lutz and Rybicki showed\cite{lutz78} that  
measurements of reactions $\pi N\to \pi^+ \pi^- N$ and $KN\to K\pi  
N$ on polarized target yield enough observables that model  
independent amplitude analysis is possible determining the spin  
dependent production amplitudes. The measurement of these reactions  
is of special interest to hadron spectroscopy because they permit to  
study the spin dependence of resonance production directly on the  
level of spin-dependent production amplitudes. Several such  
measurements were done at CERN-PS.

The high statistics measurement of $\pi^- p \to\pi^-\pi^+ n$ at  
17.2 GeV/c on unpolarized target\cite{grayer74} was later repeated  
with a transversely polarized target at the same  
energy\cite{groot78,becker79,becker79b,chabaud83,sakrej84,rybicki85}.  
Model independent amplitude analyses were performed for various  
intervals of dimeson mass at small momentum transfers $-t =  
0.005-0.2$ (GeV/c)$^2$\cite{groot78,becker79,becker79b,chabaud83},  
and over a large interval of momentum transfer $-t = 0.2-1.0$  
(GeV/c)$^2$\cite{sakrej84,rybicki85}.

Additional information was provided by the first measurement of  
$\pi^+ n \to\pi^+\pi^- p$ and $K^+ n\to K^+ \pi^- p$ reactions on  
polarized deuteron target at 5.98 and 11.85  
GeV/c\cite{lesquen85,lesquen89}. The data allowed to study the  
$t$-evolution of mass dependence of moduli of  
amplitudes\cite{svec90}. Detailed amplitude  
analyses\cite{svec92,svec92b} determined the mass dependence of  
amplitudes at larger momentum transfers $-t = 0.2-0.4$ (GeV/c)$^2$.

The crucial finding of all these measurements was the evidence for  
strong dependence of production amplitudes on nucleon spin. The  
process of resonance production is very closely related to nucleon   
transversity, or nucleon spin component in direction perpendicular  
to the production plane. For instance, in $\pi^- p\to\pi^-\pi^+ n$  
at small $t$ and dipion masses below 1000 MeV, all amplitudes with  
recoil nucleon transversity ``down'' are smaller than transversity  
``up'' amplitudes, irrespective of dimeson spin and helicity. In  
particular, the $S$-wave amplitude with recoil nucleon transversity  
``up'' is found to resonate at 750 MeV in both  
solutions\cite{svec92c,svec96,svec96b} irrespective of the method of  
amplitude analysis\cite{svec96b}, while the $S$-wave amplitude with  
recoil nucleon transversity ``down'' is nonresonating and large in  
both solutions. It is important to stress that the discovery of the  
narrow scalar state $\sigma(750)$ in $\pi^- p \to \pi^-\pi^+ n$ and  
$\pi^+ n \to \pi^+ \pi^- p$\cite{svec96,svec96b} was possible only  
because these reactions were measured on polarized targets which  
allowed the model-independent determination of the spin dependent  
production amplitudes.

The assumption that production amplitudes in $\pi^- p\to\eta\pi^-  
p$ and $\pi^- p\to\eta\pi^0 n$ do not depend on nucleon spin  
contradicts all that we have learned from the measurements of $\pi  
N\to \pi^+\pi^- N$ on polarized targets at CERN. Applied to  
reactions $\pi^- p\to\pi^- \pi^+ n$ and $\pi^+ n\to\pi^+ \pi^- p$,  
the assumption has observable consequences that can be tested  
directly in measurements on polarized targets. In the previous  
paper\cite{svec96c} we have shown how all these consequences are in  
contradiction with CERN polarized data on $\pi N_\uparrow  
\to\pi^+\pi^- N$ and $K^+ n_\uparrow \to K^+ \pi^- p$ (see Fig.~1  
and 2 of Ref.~26). We must conclude that the CERN polarized data  
invalidate the assumption that production amplitudes do not depend  
on nucleon spin. Consequently, some of the results of analyses of  
$\pi^- p\to\eta\pi^- p$ and $\pi^- p\to\eta\pi^0 n$ may not be  
reliable.

The question of reliability of amplitude analyses based on  
assumption of independence of production amplitudes on nucleon spin  
is of special importance to searches for exotic resonances like  
$1^{-+}(1405)$ in $\pi^- p\to\eta\pi^- p$ and $\pi^- p\to\eta\pi^0  
n$ reactions, or confirmation of the narrow $\sigma(750)$ state in  
$\pi^- p \to\pi^0\pi^0 n$ reaction.

Only a model independent analysis will resolve questions concerning  
the existence of such resonances which are not seen in the  
integrated mass spectrum but only on the level of spin dependent  
production amplitudes.

In the previous paper\cite{svec96c} we have shown how measurements  
of $\pi^- p\to\pi^0\pi^0 n$ on polarized targets allow a model  
independent amplitude analysis of this reaction (and $\pi^-  
p\to\eta\eta n$). Using the results of Lutz and  
Rybicki\cite{lutz78}, we show in this work that measurements of  
$\pi^- p\to\eta\pi^- p$ and $\pi^- p\to\eta\pi^0 n$ on polarized  
target again allow a model independent determination of moduli of  
all production amplitudes and cosines of certain independent  
relative phases. We find an eight-fold ambiguity, which is the same  
situation as in model dependent analyses of unpolarized data.

We propose that high statistics measurements of $\pi^-  
p\to\eta\pi^- p$ and $\pi^- p\to\eta\pi^0 n$ be made at Brookhaven  
Multiparticle Spectrometer and at IHEP Protvino in conjunction with  
measurements of $\pi^- p\to\pi^0\pi^0$ reaction on polarized target.

The paper is organized as follows. In Section II we review our  
basic notation and definitions of observables and amplitudes. In  
Section III we present the expressions for unpolarized and polarized  
moments in terms of amplitudes. In Section IV we discuss the method  
of model-independent amplitude analysis of data on $\pi^-  
p\to\eta\pi^- p$ and $\pi^- p\to\eta\pi^0 n$ on polarized target.  
The paper closes with Section V where we present summary and our  
proposals.

\section{Basic Formalism.}

The kinematical variables which describe the reactions $\pi^-  
p\to\eta\pi^- p$ and $\pi^- p \to\eta\pi^0 n$ on a polarized proton  
target at rest are $(s,t,m,\theta,\phi, \psi, \delta$) where $s$ is  
the c.m.s. energy squared, $t$ is four-momentum transfer to the  
nucleon squared, and $m$ is the invariant mass of the $\eta\pi$  
system. The angles $\theta, \phi$ describe the direction of $\eta$  
in the $\eta\pi^-$ or $\eta\pi^0$ rest frame. The angle $\psi$ is  
the angle between the direction of target transverse polarization  
and the normal ${\vec n}$ to the scattering plane (Fig.~1). The  
direction of normal ${\vec n}$ is defined according to Basel  
convention by ${\vec p}_\pi \times {\vec p}_{\eta\pi}$ where ${\vec  
p}_\pi$ and ${\vec p}_{\eta\pi}$ are the incident and dimeson  
momenta in the target proton rest frame. The angle $\delta$ is the  
angle between the direction of target polarization vector and its  
transverse component (Fig.~1). The analysis is usually carried out  
in the $t$-channel helicity frame for the $\eta\pi$ dimeson system.  
The helicities of initial and final nucleons are always defined in  
the $s$-channel helicity frame.

When the polarization of the recoil nucleon is not measured, the  
unnormalized angular distribution of $\eta\pi^-$ or $\eta\pi^0$  
production on polarized protons at rest at fixed $s$, $m$ and $t$  
can be written\cite{lutz78} as

\begin{equation}
I(\Omega,\psi,\delta) = I_U (\Omega) + P_T \cos\psi I_C (\Omega) +  
P_T \sin\psi I_S (\Omega) + P_L I_L (\Omega)
\end{equation}

\noindent
where $P_T = P\cos\delta$ and $P_L = P\sin\delta$ are the  
transverse and longitudinal components of target polarization ${\vec  
P}$ with respect to the incident momentum (Fig.~1). In the data  
analysis of angular distribution of the dimeson system, it is  
convenient to use expansions of the angular distributions in terms  
of spherical harmonics. In the notation of Lutz and  
Rybicki\cite{lutz78} we have

\begin{equation}
I_U (\Omega) = \sum\limits_{L,M} t^L_M {\rm Re} Y^L_M (\Omega)
\end{equation}

\[
I_C(\Omega) = \sum\limits_{L,M} p^L_M {\rm Re} Y^L_M (\Omega)
\]

\[
I_S(\Omega) = \sum\limits_{L,M} r^L_M {\rm Im} Y^L_M (\Omega)
\]

\[
I_L(\Omega) = \sum\limits_{L,M} q^L_M {\rm Im} Y^L_M (\Omega)
\]

\noindent
The moments $t^L_M$ are unpolarized and are measured in experiments  
on unpolarized targets. Experiments with transversely polarized  
targets measure transverse polarized moments $p^L_M$ and $r^L_M$ but  
not the longitudinal polarized moments $q^L_M$. More details on  
these observables are given in Ref.~10 and 26.

The reaction $\pi^- p\to\eta\pi^- p$ (or $\pi^- p\to\eta\pi^0 n$)  
is described by production amplitude $H_{\lambda_n,0\lambda_p}  
(s,t,m,\theta,\phi)$ where $\lambda_p$ and $\lambda_n$ are the  
helicities of the proton and neutron, respectively. The production  
amplitudes can be expressed in terms of production amplitudes  
corresponding to definite dimeson spin $J$ and helicity $\lambda$  
using an angular expansion

\begin{equation}
H_{\lambda_n, 0\lambda_p} = \sum\limits^\infty_{J=0}  
\sum\limits^{+J}_{\lambda = -J} (2J+1)^{1/2} H^J_{\lambda\lambda_n,  
0\lambda_p} (s,t,m) d^J_{\lambda 0} (\theta) e^{i\lambda\phi}
\end{equation}

\noindent
In the following we will consider only $S$-wave $(J=0)$, $P$-wave  
$(J=1)$ and $D$-wave $(J=2)$ amplitudes. Since the experimental  
moments with $M>2$ vanish, we will restrict the dimeson helicity  
$\lambda$ only to values $\lambda = 0$ and $\lambda = \pm 1$.

The amplitudes $H^J_{\lambda\lambda_n,0\lambda_p} (s,t,m)$ can be  
expressed in terms of nucleon helicity amplitudes with definite  
$t$-channel exchange naturality. The nucleon $s$-channel helicity  
amplitudes describing the production of $\eta\pi^-$ (or $\eta\pi^0$)  
system in the $S$-, $P$- and $D$-wave states are:

\begin{equation}
0^- {1\over 2}^+ \to 0^+ {1\over 2}^+\ \colon\ H^0_{0+,0+}  = S_0,\  
H^0_{0+,0-} = S_1
\end{equation}

\[
0^- {1\over 2}^+ \to 1^- {1\over 2}^+\ \colon\ H^1_{0+,0+}  =  
P^0_0,\ H^1_{0+,0-} = P^0_1
\]

\[
H^1_{\pm 1 +,0+} = {{P^+_0 \pm P^-_0}\over{\sqrt 2}},\  
H^1_{\pm1+,0-} = {{P^+_1 \pm P^-_1}\over{\sqrt 2}}
\]

\[
0^- {1\over 2}^+ \to 2^+ {1\over 2}^+\ \colon\ H^2_{0+,0+} =  
D^0_0,\ H^2_{0+,0-} = D^0_1
\]

\[
H^2_{\pm 1+,0+} = {{D^+_0 \pm D^-_0}\over{\sqrt 2}},\ H^2_{\pm  
1+,0-} = {{D^+_1 \pm D^+_1}\over{\sqrt 2}}
\]

\noindent
At large $s$, the amplitudes $S_n, P^0_n, P^-_n, D^0_n, D^-_n,  
n=0,1$ are dominated by the unnatural exchanges. The amplitudes  
$P^+_n, D^+_n, n=0,1$ are dominated by natural exchanges. The index  
$n=|\lambda_p - \lambda_n|$ is nucleon helicity flip.

The observables measured in experiments on transversely polarized  
targets are most simply related to nucleon transversity amplitudes  
of definite naturality\cite{lutz78,lesquen89,kotanski66}. With  
$k=1/\sqrt 2$, they are defined as follows:

\begin{equation}
S = k(S_0 + iS_1)\ ,\ \overline S = k(S_0 - iS_1)
\end{equation}

\[
P^0 = k(P^0_0 + iP^0_1)\ ,\ \overline P^0 = k(P^0_0 - iP^0_1)
\]

\[
 P^- = k(P^-_0 + iP^-_1)\ ,\ \overline P^-  = k (P^-_0 - iP^-_1)
\]

\[
P^+ = k(P^+_0 - iP^+_1)\ ,\ \overline P^+ = k(P^+_0 + iP^+_1)
\]

\[
D^0 = k(D^0_0 + iD^0_1)\ ,\ \overline D^0 = k(D^0_0 - iD^0_1)
\]

\[
D^- = k(D^-_0 + iD^-_1)\ ,\ \overline D^- = k(D^-_0 - iD^-_1)
\]

\[
D^+ = k(D^+_0 - iD^+_1)\ ,\ \overline D^+ = k(D^+_0 + iD^+_1)
\]

\noindent
The nucleon helicity and nucleon transversity amplitudes differ in  
the quantization axis for the nucleon spin. The transversity  
amplitudes $S, P^0, P^-, P^+, D^0, D^-, D^+$ $(\overline S,  
\overline P^0, \overline P^-, \overline P^+, \overline D^0,  
\overline D^-, \overline D^+)$ describe the production of $\eta\pi$  
state with the recoil nucleon spin antiparallel or down (parallel or  
up) relative to the normal ${\vec n}$ to the production plane.

\section{Observables in terms of amplitudes.}

It is useful to express the moments $t^L_M$ and $p^L_M$ in terms of  
quantities that do not depend explicitly on whether we use nucleon  
helicity or nucleon transversity amplitudes. The required quantities  
are spin-averaged partial wave intensity

\begin{equation}
I_A = |A|^2 + |\overline A|^2 = |A_0|^2 + |A_1|^2
\end{equation}

\noindent
and partial wave polarization

\begin{equation}
P_A = |A|^2 - |\overline A|^2 = 2\epsilon_A {\rm Im} (A_0 A^*_1)
\end{equation}

\noindent
where $\epsilon_A = +1$ for $A = S, P^0, P^-, D^0, D^-$ and  
$\epsilon_A = -1$ for $A = P^+, D^+$. We also need spin-averaged  
interference terms

\begin{equation}
R(AB) = {\rm Re} (AB^* + \overline A\; \overline B^*) = {\rm Re}  
(A_0 B_0 + \epsilon_A \epsilon_B A_1 B_1)
\end{equation}

\begin{equation}
Q(AB) = {\rm Re} (AB^* - \overline A\; \overline B^*) = {\rm Re}  
(\epsilon_B A_ 0 B^*_1 - \epsilon_A A_1 B^*_0)
\end{equation}

\noindent
Then moments $t^L_M$ are expressed in terms of intensities $I_A$  
and interference terms $R(AB)$. The moments $p^L_M$ are expressed in  
terms of polarizations $P_A$ and interference terms $Q(AB)$. The  
moments $r^L_M$ are interferences between the natural and unnatural  
exchange amplitudes. To describe moments $r^L_M$, it is useful to  
introduce notation

\begin{equation}
N(AP^+) = {\rm Re} (AP^{+*} - \overline A\; \overline P^{+*})
\end{equation}

\[
N(AD^+) = {\rm Re} (AD^{+*} - \overline A\; \overline D^{+*})
\]

\noindent
where $A = S, P^0, P^-, D^0, D^-$.

Using the results of the Lutz and Rybicki\cite{lutz78} we obtain  
the following expressions for moments with $c=\sqrt{4\pi}$:

\FL
\begin{equation}
{\rm Unpolarized\  moments}\  t^L_M
\end{equation}

\[
ct^0_0 = I_S + I_{P^0} + I_{P^-} + I_{P^+} + I_{D^0} + I_{D^-} + I_{D^+}
\]

\[
ct^1_0 = 2R (SP^0) + {4\over\sqrt 5} R (P^0 D^0) + 2\sqrt{{3\over  
5}} [R (P^- D^-) + R(P^+ D^+)]
\]

\[
ct^1_1 = 2\sqrt 2 R(SP^-) + 2\sqrt{{6\over 5}} R(P^0 D^-) -  
2\sqrt{{2\over 5}} R(P^- D^0)
\]

\[
ct^2_0 = {2\over\sqrt 5} I_{P^0} - {1\over\sqrt 5} (I_{P^-} +  
I_{P^+} ) + 2R (SD^0) + {2\over 7} \sqrt 5 I_{D^0} + {\sqrt 5\over  
7} (I_{D^-} + I_{D^+})
\]

\[
ct^2_1 = 2\sqrt{{6\over 5}} R(P^0P^-) + 2\sqrt 2 R(SD^-) +  
{{2\sqrt{10}}\over 7} R(D^0 D^-)
\]

\[
ct^2_2 = \sqrt{{6\over 5}} (I_{P^-} - I_{P^+}) + {{\sqrt{30}}\over  
7} (I_{D^-} - I_{D^+})
\]

\[
ct^3_0 = 6\sqrt{{3\over{35}}} R(P^0 D^0) - {6\over{\sqrt{35}}}  
[R(P^- D^-) + R(P^+ D^+)]
\]

\[
ct^3_1 = 8 \sqrt{{3\over{35}}} R(P^0  D^-) + {{12}\over{\sqrt{35}}}  
R(P^- D^0)
\]

\[
ct^3_2 = 2\sqrt{{6\over 7}} [R(P^- D^-) - R(P^+ D^+)]
\]

\[
ct^4_0 = {6\over 7} I_{D^0} - {4\over 7} (I_{D^-} + I_{D^+})
\]

\[
ct^4_1 = {4\over 7} \sqrt{15} R (D^0 D^-)
\]

\[
ct^4_2 = {{2\sqrt{10}}\over{7}} (I_{D^-} - I_{D^+})
\]

\FL
\begin{equation}
{\rm Polarized\ moments}\ p^L_M
\end{equation}

\[
cp^0_0 = P_S + P_{P^0} + P_{P^-} - P_{P^+} + P_{D^0} + P_{D^-} - P_{D^+}
\]

\[
cp^1_0 = 2Q (SP^0) + {4\over{\sqrt 5}} Q(P^0 D^0) + 2 \sqrt{{3\over  
5}} [Q(P^- D^-) - Q(P^+ D^+)]
\]

\[
cp^1_1 = 2\sqrt 2 Q (SP^-) + 2 \sqrt{{6\over 5}} Q(P^0 D^-) - 2  
\sqrt{{2\over 5}} Q(P^- D^0)
\]

\[
cp^2_0 = {2\over\sqrt 5} P_{P^0} - {1\over{\sqrt 5}} (P_{P^-} -  
P_{P^+}) + 2Q (SD^0) + {{2\sqrt 5}\over 7} P_{D^0} + {{\sqrt 5}\over  
7} (P_{D^-} - P_{D^+})
\]

\[
cp^2_1 = 2 \sqrt{{6\over 5}} Q(P^0P^-) + 2\sqrt 2 Q (SD^-) +  
{{2\sqrt{10}}\over 7} Q(D^0 D^-)
\]

\[
cp^2_2 = \sqrt{{6\over 5}} (P_{P^-} + P_{P^+}) + {{\sqrt{30}}\over  
7} (P_{D^-} + P_{D^+})
\]

\[
cp^3_0 = 6 \sqrt{{3\over{35}}} Q (P^0 D^0) - {6\over{\sqrt{35}}} [Q  
(P^- D^-) - Q (P^+ D^+)]
\]

\[
cp^3_1 = 8 \sqrt{{3\over{35}}} Q (P^0 D^-) + {{12}\over{\sqrt{35}}}  
Q(P^- D^0)
\]

\[
cp^3_2 = 2\sqrt{{6\over 7}} [Q (P^- D^-) + Q (P^+ D^+)]
\]

\[
cp^4_0 = {6\over 7} P_{D^0} - {4\over 7} (P_{D^-} - P_{D^+})
\]

\[
cp^4_0 = {4\over 7} \sqrt{15} Q(D^0 D^-)
\]

\[
cp^4_2 = {{2\sqrt{10}}\over 7} (P_{D^-} + P_{D^+})
\]

\FL
\begin{equation}
{\rm Polarized\ moments}\ r^L_M
\end{equation}

\[
cr^1_1 = -2\sqrt 2 N(SP^+) - 2\sqrt{{2\over 5}} N(D^0 P^+) - 2  
\sqrt{{6\over 5}} N(P^0 D^+)
\]

\[
cr^2_1 = -2 \sqrt{{6\over 5}} N(P^0 P^+) - 2\sqrt 2 N(SD^+) -  
{{2\sqrt{10}}\over 7} N(D^0 D^+)
\]

\[
cr^2_2 = - 2 \sqrt{{6\over 5}} N(P^- P^+) - {{2\sqrt{30}}\over 7}  
N(D^- D^+)
\]

\[
cr^3_1 = + {{12}\over{\sqrt{35}}} N(D^0 P^+) - 8  
\sqrt{{3\over{35}}} N(P^0 D^+)
\]

\[
cr^3_2 = - 2 \sqrt{{6\over 7}} N(D^- P^+) -2 \sqrt{{6\over 7}} N(P^- D^+)
\]

\[
cr^4_1 = - {4\over 7} \sqrt{15} N(D^0 D^+)
\]

\[
cr^4_2 = - {4\over 7} \sqrt{10} N(D^- D^+)
\]

\section{Model independent amplitude analysis.}

Our starting point is the observation of symmetry in the relations  
for moments $t^L_M$ and $p^L_M$. We find that we get $p^L_M$ from  
$t^L_M$ by replacing intensities $I_A$ by polarizations $\epsilon_A  
P_A$, $\epsilon = +1$ for $A = S, P^0, P^-, D^0, D^-$ and  
$\epsilon_A = -1$ for $A = P^+, D^+$, and by replacing the  
interference terms $R(AB) \to Q(AB)$ for unnatural exchange  
amplitudes and $R(P^+ D^+)\to - Q (P^+D^+)$ for natural exchange  
amplitudes. To solve the system of equations $t^L_M$ and $p^L_M$ it  
will be useful to work with transversity amplitudes. Then the  
definitions (3.1)--(3.4) suggest to construct two sets of equations  
corresponding to the sum and difference of the moments $t^L_M$ and  
$p^L_M$. In this way we get two independent sets of equations for  
amplitudes of opposite transversity.

The first set of new observables reads:

\begin{equation}
a_1 = {c\over 2} (t^0_0 + p^0_0) = |S|^2 + |P^0|^2 + |P^-|^2 +  
|\overline P^+|^2 + |D^0|^2 + |D^-|^2 + |\overline D^+|^2
\end{equation}

\[
a_2 = {c\over 2} (t^1_0 + p^1_0) = 2{\rm Re} (SP^{0*}) +  
{4\over{\sqrt 5}} {\rm Re} (P^0 D^{0*}) + 2 \sqrt{{3\over 5}} [{\rm  
Re} (P^- D^{-*}) + {\rm Re} (\overline P^+ \overline D^{+*})]
\]

\[
a_3 = {c\over 2} (t^1_1 + p^1_1) = 2\sqrt 2 {\rm Re} (SP^{-*}) +  
2\sqrt{{6\over 5}} {\rm Re} (P^0 D^{-*}) - 2 \sqrt{{2\over 5}} {\rm  
Re} (P^- D^{0*})
\]

\[
a_4 = {c\over 2} (t^2_0 + p^2_0) = {2\over\sqrt 5} |P^0|^2 -  
{1\over{\sqrt 5}} (|P^-|^2 + |\overline P^+|^2) + 2 {\rm Re}  
(SD^{0*}) + {{2\sqrt 5}\over 7} |D^0|^2 + {\sqrt 5\over 7} (|D^-|^2  
+ |\overline D^+|^2)
\]

\[
a_5 = {c\over 2} (t^2_1 + p^2_1) = 2 \sqrt{{6\over 5}} {\rm Re}  
(P^0 P^{-*}) + 2\sqrt 2 {\rm Re} (SD^{-*}) + {{2\sqrt{10}}\over 7}  
{\rm Re} (D^0 D^{-*})
\]

\[
a_6 = {c\over 2} (t^2_2 + p^2_2) = \sqrt{{6\over 5}} (|P^-|^2 -  
|\overline P^+|^2) + {{\sqrt{30}}\over 7} (|D^-|^2 - |\overline  
D^+|^2)
\]

\[
a_7 = {c\over 2} (t^3_0 + p^3_0) = 6 \sqrt{{3\over{35}}} {\rm Re}  
(P^0 D^{0*}) - {6\over{\sqrt{35}}} [{\rm Re} (P^- D^{-*}) + {\rm Re}  
(\overline P^+ \overline D^{+*})]
\]

\[
a_8 = {c\over 2} (t^3_1 + p^3_1) = 8\sqrt{{3\over{35}}} {\rm Re}  
(P^0 D^{-*}) + {{12}\over{\sqrt{35}}} {\rm Re} (P^- D^{0*})
\]

\[
a_9 = {c\over 2} (t^3_2 + p^3_2) = 2\sqrt{{6\over 7}} [{\rm Re}  
(P^- D^{-*}) - {\rm Re} (\overline P^+ \overline D^{+*})]
\]

\[
a_{10} = {c\over 2} (t^4_0 + p^4_0) = {6\over 7} |D^0|^2 - {4\over  
7} (|D^-|^2 + |\overline D^+|^2)
\]

\[
a_{11} = {c\over 2} (t^4_1 + p^4_1) = {4\over 7} \sqrt{15} {\rm Re}  
(D^0 D^{-*})
\]

\[
a_{12} = {c\over 2} (t^4_2 + p^4_2) = {{2\sqrt{10}}\over 7}  
(|D^-|^2 - |\overline D^+|^2)
\]

\noindent
The first set of equations (4.1) involves 7 moduli

\begin{equation}
|S|, |P^0|, |P^-|, |\overline P^+|, |D^0|, |D^-|, |\overline D^+|
\end{equation}

\noindent
and 10 cosines of relative phases between unnatural exchange amplitudes

\begin{equation}
\cos(\gamma_{SP^0}), \cos(\gamma_{SP^-}), \cos(\gamma_{SD^0}),  
\cos(\gamma_{SD^-})
\end{equation}

\begin{equation}
\cos(\gamma_{P^0P^-}), \cos(\gamma_{P^0D^0}), \cos(\gamma_{P^0D^-})
\end{equation}

\begin{equation}
\cos(\gamma_{P^-D^0}), \cos(\gamma_{P^- D^-}), \cos(\gamma_{D^0D^-})
\end{equation}

\noindent
and 1 cosine of relative phase between natural exchange amplitudes

\begin{equation}
\cos(\overline\gamma_{P^+D^+})
\end{equation}

The second set of observables $\overline a_i, i=1,2,\ldots,12$  
corresponding to the differences of moments $t^L_M$ and $p^L_M$  
involves the same moduli and cosines as the first set but for  
amplitudes of opposite transversity:

\noindent
7 moduli

\begin{equation}
|\overline S|, |\overline P^0|, |\overline P^-|, |P^+|, |\overline  
D^0|, |\overline D^-|, |D^+|
\end{equation}

\noindent
10 cosines of relative phases between unnatural exchange amplitudes

\begin{equation}
\cos (\overline\gamma_{SP^0}), \cos(\overline\gamma_{SP^-}),  
\cos(\overline\gamma_{SD^0}), \cos(\overline\gamma_{SD^-})
\end{equation}

\[
\cos(\overline\gamma_{P^0P^-}), \cos(\overline\gamma_{P^0D^0}),  
\cos(\overline\gamma_{P^0D^-})
\]

\begin{equation}
\cos(\overline\gamma_{P^-D^0}), \cos(\overline\gamma_{P^-D^-}),  
\cos(\overline\gamma_{D^0D^-})
\end{equation}

\noindent
1 cosine of relative phase between natural exchange amplitudes

\begin{equation}
\cos(\gamma_{P^+D^+})
\end{equation}

We will now show that the cosines (4.4) and (4.5) can be expressed  
in terms of the cosines (4.3). For instance, we can write

\begin{equation}
\gamma_{P^0P^-} = \phi_{P^0} - \phi_{P^-} = (\phi_S - \phi_{P^-}) -  
(\phi_S - \phi_{P^0}) = \gamma_{SP^-} - \gamma_{SP^0}
\end{equation}

\noindent
Then

\begin{equation}
\cos(\gamma_{P^0P^-}) = \cos(\gamma_{SP^0}) \cos(\gamma_{SP^-}) +  
\sin (\gamma_{SP^0}) \sin(\gamma_{SP^-})
\end{equation}

\noindent
Since the signs of the sines $\sin(\gamma_{SP^0})$ and  
$\sin(\gamma_{SP^-})$ are not known, we write

\begin{equation}
\sin(\gamma_{SP^0}) = \epsilon_{SP^0} |\sin(\gamma_{SP^0})|
\end{equation}

\[
\sin(\gamma_{SP^-}) = \epsilon_{SP^-} |\sin (\gamma_{SP^-})|
\]

\noindent
Hence

\begin{equation}
\cos(\gamma_{P^0P^-}) = \cos(\gamma_{SP^0}) \cos(\gamma_{SP^-}) +  
\epsilon_{P^0P^-} \sqrt{(1-\cos^2 \gamma_{SP^0})  
(1-\cos^2\gamma_{SP^0})}
\end{equation}

\noindent
where $\epsilon_{P^0P^-} = \pm 1$ is the sign ambiguity. The  
remaining cosines in (4.4) and (4.5) can be written in the form  
similar to (4.14) with their own sign ambiguities. The sign  
ambiguities of cosines (4.4) and (4.5) can be expressed in terms of  
sign ambiguities corresponding to the sines $\sin(\gamma_{SP^0})$,  
$\sin(\gamma_{SP^-})$, $\sin(\gamma_{SD^0})$ and  
$\sin(\gamma_{SD^-})$. We can write

\begin{equation}
\epsilon_{P^0P^-} = \epsilon_{SP^0} \epsilon_{SP^-}
\end{equation}

\[
\epsilon_{P^0D^0} = \epsilon_{SP^0} \epsilon_{SD^0}
\]

\[
\epsilon_{P^0D^-} = \epsilon_{SP^0} \epsilon_{SD^-}
\]

\begin{equation}
\epsilon_{P^-D^0} = \epsilon_{SP^-} \epsilon_{SD^0}
\end{equation}

\[
\epsilon_{P^-D^-} = \epsilon_{SP^-} \epsilon_{SD^-}
\]

\[
\epsilon_{D^0D^-} = \epsilon_{SD^0} \epsilon_{SD^-}
\]

\noindent
The reversal of all signs $\epsilon_{SP^0}$, $\epsilon_{SP^-}$,  
$\epsilon_{SD^0}$ and $\epsilon_{SD^-}$ yields the same signs in  
(4.15) and (4.16). The sign ambiguities (4.16) are not independent.  
They are uniquely determined by the sign ambiguities (4.15). Only  
sign ambiguities (4.15) are independent and there is 8 sign  
combinations in (4.15). The following table lists all eight allowed  
sets of sign ambiguities of cosines (4.4) and (4.5):

\newpage

\begin{quasitable}
\begin{tabular}{lcccccccc}
&1 &2 &3 &4 &5 &5 &7 &8 \\
\tableline
$\epsilon_{P^0P^-}$ &$+$ &$-$ &$+$ &$+$ &$-$ &$-$ &$+$ &$-$ \\
$\epsilon_{P^0D^0}$ &$+$ &$+$ &$-$ &$+$ &$-$ &$+$ &$-$ &$-$ \\
$\epsilon_{P^0D^-}$ &$+$ &$+$ &$+$ &$-$ &$+$ &$-$ &$-$ &$-$ \\
$\epsilon_{P^-D^0}$ &$+$ &$-$ &$-$ &$+$ &$+$ &$-$ &$-$ &$+$ \\
$\epsilon_{P^-D^-}$ &$+$ &$-$ &$+$ &$-$ &$-$ &$+$ &$-$ &$+$ \\
$\epsilon_{D^0D^-}$ &$+$ &$+$ &$-$ &$-$ &$-$ &$-$ &$+$ &$+$
\end{tabular}
\end{quasitable}

Using expressions like (4.14) for cosines (4.4) and (4.5), the  
number of unknowns is reduced to 12. With each choice of sign  
ambiguity from the above Table we have a set of 12 equations for 12  
unknown which can be solved numerically or by $\chi^2$ method. Of  
course, there is an eight-fold ambiguity and we obtain 8 solutions  
for moduli (4.2) and cosines (4.3) and (4.6) in each $(m,t)$ bin.  
Since each solution is uniquely labeled by the choice of sign  
ambiguities, there is no problem linking solutions in neighbouring  
$(m,t)$ bins. Similarly we obtain 8 solutions for moduli (4.7),  
cosines (4.8) and (4.11) from the second set of equations,  
$\overline a_i, i=1,2,\ldots,12$.

The 8 solutions from the first set of equations $a_i,  
i=1,2,\ldots,12$ are independent from the 8 solutions obtained from  
the second set of equations $\overline a_i, i=1,2,\ldots,12$.  
Consequently, there will be a 64-fold ambiguity in the partial wave  
intensities which we can write

\begin{equation}
I_A (i,j) = |A(i)|^2 + |\overline A (j)|^2, i,j=1,2,\ldots 8
\end{equation}

\noindent
where $A = S, P^0, P^-, P^+, D^0, D^-, D^+$.

As in the case of amplitude analysis of $\pi^- p_\uparrow \to \pi^+  
\pi^+ n$ at 17.2  
GeV/c\cite{groot78,becker79,becker79b,chabaud83,sakrej84,rybicki85},  
the unpolarized moments $t^L_M$ should come from measurements on  
unpolarized targets.

\section{Summary.}

Measurements of $\pi^- p \to \pi^-\pi^+ n$, $\pi^+ n \to \pi^+  
\pi^- p$ and $K^+ n \to K^+ \pi^- p$ on polarized targets at CERN  
found evidence for a strong dependence of pion production amplitudes  
on nucleon spin. This evidence invalidates the  
assumption\cite{costa80,chung96} that production amplitudes in  
$\pi^- p\to\eta\pi^- p$ and $\pi^- p\to\eta\pi^0 n$ reactions do not  
depend on nucleon spin. The amplitude analyses of these reactions  
based on the assumption of independence of production amplitudes on  
nucleon spin are thus insufficient and are likely to be unreliable.  
To ascertain the existence of exotic resonance $1^{-+} (1405)$ and  
study its properties, a reliable, model independent amplitude  
analysis is required. Nucleon spin is not only relevant to the  
dynamics of production processes. It also allows the model  
independent determination of spin-dependent production amplitudes  
from measurements of $\pi^- p \to \eta\pi^- p$ and $\pi^-  
p\to\eta\pi^0 n$ on polarized targets, as we have shown. Our major  
assumption was that moments with $M>2$ do not contribute to the  
angular distributions. This may not be true at large momentum  
transfers. In this case one has to use the formalism developed by  
I.~Sakrejda\cite{sakrej84} which takes into account the helicities  
$\lambda=\pm 2$ of the $D$-wave.

Instruments shape research and determine which discoveries are  
made. Polarized targets have proven themselves to be valuable and  
important tools of discovery. We propose that high statistics  
measurements of reactions $\pi^- p\to\eta\pi^- p$ and $\pi^-  
p\to\eta\pi^0 n$ be made on polarized targets at BNL Multiparticle  
Spectrometer and at IHEP in Protvino, in conjunction with high  
statistics measurements of $\pi^- p\to\pi^0 \pi^0$ on polarized  
targets. Such experiments will be also feasible at the recently  
proposed Japanese Hadron Project (JHP). When built, JHP will be a  
high-intensity 50 GeV proton accelerator complex with high quality  
pion, kaon and antiproton secondary beams\cite{jhpsite}. The  
availability of such secondary beams will make JHP an ideal facility  
for hadron spectroscopy using polarized targets in a search for new  
resonant states at the level of spin-dependent production  
amplitudes.

\acknowledgements
I wish to thank B.B.~Brabson for triggering my interest in BNL  
E-852 measurements of $\pi^- p\to\eta \pi^- p$ and $\pi^- p  
\to\eta\pi^0 n$ reactions and to Yu.~D.~Prokoshkin for stimulating  
e-mail correspondence. This work was supported by Fonds pour la  
Formation de Chercheurs et l'Aide \`a la Recherche (FCAR),  
Minist\`ere de l'Education du Qu\'ebec, Canada.

\begin{figure}
\caption{Definition of the coordinate system used to describe the  
target polarization ${\vec P}$ and the decay of the dimeson  
$\eta\pi^-$ system.}\label{fig1}
\end{figure}

\end{document}